\documentclass[aps,pra,showpacs,groupedaddress,floatfix,nofootinbib]{revtex4}
\usepackage{amsmath}
\usepackage{graphicx}
\usepackage{color}
\newcommand{\beq}{\begin{eqnarray}}
\newcommand{\eeq}{\end{eqnarray}}

\begin{document}

\title{{Variational collocation on finite intervals}}
\author{Paolo Amore}
\email{paolo@ucol.mx}
\affiliation{Facultad de Ciencias, Universidad de Colima,\\
and Centro Universitario de Ciencias Basicas, \\
Bernal D\'{i}az del Castillo 340, Colima, Colima,\\
Mexico.}
\author{Mayra Cervantes}
\affiliation{Facultad de Ciencias, Universidad de Colima,\\
and Centro Universitario de Ciencias Basicas, \\
Bernal D\'{i}az del Castillo 340, Colima, Colima,\\
Mexico.}

\author{Francisco M. Fern\'andez}\email{fernande@quimica.unlp.edu.ar}
\affiliation{INIFTA (Conicet,UNLP), Diag. 113 y 64 S/N, \\
Sucursal 4, Casilla de Correo 16, 1900 La Plata, Argentina}

\begin{abstract}
In this paper we study a new family of sinc--like functions,
defined on an interval of finite width. These functions, which we
call ``little sinc'', are orthogonal and share many of the
properties of the sinc functions. We show that the little sinc
functions supplemented with a variational approach enable one to
obtain accurate results for a variety of problems. We apply them
to the interpolation of functions on finite domain and to the
solution of the Schr\"odinger equation, and compare the
performance of present approach with others.
\end{abstract}


\maketitle

\section{Introduction}
\label{intro}

The main goal of this paper is to introduce a new set of
orthogonal functions, hereafter called "little sinc functions"
(LSF), and show that they can be used to solve a wide class of
problems, such as function interpolation, the numerical solution
of differential equations, including the Schr\"odinger equation.
Actually the LSF, which are defined on a finite domain, were first
derived by D. Baye~\cite{Baye95} in the framework of generalized
meshes, but the relation with the usual sinc functions was not
recognized. As a matter of fact, in recent years Baye and
collaborators have used these and other functions for the solution
of the Schr\"odinger equation with several different potentials,
and they have produced accurate numerical results both for the
energies and wave functions~\cite{Baye95,Baye86,BGV02}.

On the other hand, sinc collocation methods have been applied to a
larger class of problems, which include those mentioned above (see
for example \cite{Kour96}), and they have been put on firm
mathematical grounds~\cite{Ste93}. Examples of applications of the
sinc functions can be found in references
\cite{Ste79,Morlet95,Lybeck96,Nara02,Eas00,Rev03}, among others.
Recently, one of the authors has shown that the sinc collocation
methods can be optimized by using a variational
approach~\cite{Amore06} based on the Principle of Minimal
Sensitivity~\cite{Ste81}. This optimization allows one to obtain
the highest precision with a given number of grid points, and can
be particularly valuable in problems that require intense
numerical calculation.

We feel that it is important to establish the relation of the LSF
with the sinc functions at least for two reasons: first, because
it will be possible to generalize the variational approach
\cite{Amore06} to the LSF, thus improving the convergence behavior
of the method, and second because it offers an interesting link to
generalized mesh methods and to open new areas of application of
the latter.

This paper is organized as follows: in Section \ref{sinc} we
describe the general properties of the usual sinc functions,
defined on the real line. In Section \ref{little} we derive an
expression for the LSF and discuss their
properties. In Section~\ref{sch} we solve the Schr\"odinger
equation with two different potentials and compare numerical results obtained
with the usual sinc functions and LSF. Finally, in \ref{concl} we draw our
conclusions.


\section{Sinc functions}
\label{sinc}

In what follows we outline some of the basic properties of the
generalized sinc functions, defined as
\beq 
S_k(h,x) &\equiv& \frac{\sin\left(\pi/h (x-k h)\right)}{\pi/h (x-k h)}  \ , 
\label{eq_1_1} 
\eeq 
where $k = 0, \pm 1, \pm 2,\dots$.
 The sinc function for a given value of the index
 $k$ is peaked at $x_k = k h$, where it equals unity, and vanish at
the other points $x_j = j h$, with $j \neq k$ and $j = 0, \pm 1,
\pm 2, \dots$.

From the integral representation
\beq 
S_k(h,x) &=& \frac{h}{2\pi} \int_{-\pi/h}^{+\pi/h} \ e^{\pm i (x- k h) t} \ dt \ , 
\label{eq_1_2} 
\eeq
we easily derive the normalization factor
 \beq 
{\cal I}_1 &\equiv& \int_{-\infty}^{+\infty} S_k(h,x) \ dx = h   \ , 
\eeq
and the orthogonality property
\beq 
{\cal I}_2 &\equiv&
\int_{-\infty}^{+\infty} S_k(h,x) \ S_l(h,x) \ dx = h \ \delta_{kl}  . 
\eeq
It is worth noticing that eq.~(\ref{eq_1_2}) defines a Dirac delta
function in the limit $h \rightarrow 0$.

A function $f(x)$ analytic on a rectangular strip centered on the
real axis can be approximated in terms of sinc functions
as~\cite{Ste93}
\beq 
f(x) \approx  \sum_{k=-\infty}^{+\infty} \ f(k h) \ S_{k}(h,x) \ , \label{eq_1_4} 
\eeq
which together with the normalization factor can be used to approximate the definite
integral \beq \int_{-\infty}^{+\infty} f(x) \ dx &\approx& h \
\sum_{k=-\infty}^{\infty} f(k,h)  . \eeq

It is not difficult to derive simple expressions of the derivatives of sinc functions
in terms of the same sinc functions:
\beq
\frac{d}{dx} S_k(h,x)  = \sum_{l=-\infty}^{\infty} \ c_{lk}^{(1)} \  S_k(h,x)  \ ,
\eeq

where
\beq
c_{kl}^{(1)} &\equiv& \int_{-\infty}^{+\infty} S_l(h,x) \  \frac{d}{dx} S_k(h,x) \ dx = \left\{
\begin{array}{c}
0 \ \ if \ \ k=l \\
\frac{1}{h} \ \frac{(-1)^{k-l}}{k-l} \ \ if \ \ k \neq l
\end{array}
\right.
\eeq

For the second derivative we have
\beq
\frac{d^2}{dx^2} S_k(h,x)  = \sum_{l=-\infty}^{\infty} \ c_{lk}^{(2)} \  S_k(h,x)  \ ,
\eeq
where
\beq
c_{lk}^{(2)} &\equiv& \int_{-\infty}^{+\infty} S_l(h,x) \  \frac{d^2}{dx^2} S_k(h,x) \ dx =
\left\{ \begin{array}{c}
- \frac{\pi^2}{3 h^2} \ \ if \ \ k=l \\
- \frac{2}{h^2} \frac{(-1)^{k-l}}{(k-l)^2} \ \ if \ \ k \neq l
\end{array}
\right.
\eeq

General expressions for higher order derivatives are also available~\cite{Rev03}:
\beq
c_{lk}^{(2r)} &\equiv&  \frac{(-1)^{l-k}}{h^{2r} (l-k)^{2r}}  \ \sum_{k=0}^{r-1} (-1)^{k+1} \ \frac{2r!}{(2k+1)!} \pi^{2k} \ (l-k)^{2k} \\
c_{ll}^{(2r)} &\equiv& \left(\frac{\pi}{h}\right)^{2r} \ \frac{(-1)^r}{2 r+1} \\
c_{lk}^{(2r+1)} &\equiv&  \frac{(-1)^{l-k}}{h^{2r+1} (l-k)^{2r+1}}  \ \sum_{k=0}^{r} (-1)^{k} \ \frac{(2r+1)!}{(2k+1)!} \pi^{2k} \ (l-k)^{2k} \\
c_{ll}^{(2r+1)} &\equiv& 0 \ ,
\eeq
with $r= 1,2,\dots$.


\section{Little sinc functions}
\label{little}

We will now derive the new set of LSF.
We consider the orthonormal basis of the wave functions of a particle in a box with
infinite walls located at $x=\pm L$:
\beq
\psi_n(x) = \frac{1}{\sqrt{L}} \ \sin \left(\frac{n\pi}{2L} (x+L)\right)
\eeq
and define
\beq
\overline{\delta}_N(x,y) = C_N \sum_{n=1}^N \psi_n(x) \psi_n(y) = \frac{C_N}{4 L} \
\left\{ \frac{\sin \left(\frac{(2 N+1) \pi  (x-y)}{4 L}\right)}{\sin \left(\frac{\pi  (x-y)}{4 L}\right) }
-(-1)^{N}  \frac{\cos\left(\frac{(2 N+1) \pi  (x+y)}{4 L}\right)}{ \cos \left(\frac{\pi  (x+y)}{4 L}\right)}
\right\} \ .
\eeq
where $C_N$ is a constant.

Because of the completeness of the basis $\left\{ \psi_n(x) \right\}$ we have
\beq
\lim_{N\rightarrow \infty} \frac{\overline{\delta}_N(x,y)}{C_N} = \delta(x-y) \ .
\eeq

For reasons that will soon become clear, we  set $C_N = \frac{2 L}{N}$ and select even values
of $N$. To simplify the notation $h \equiv 2L/N$ will denote the grid spacing, and
$y_k \equiv \frac{2 k L}{N} = k h$, with $k = - N/2+1, -N/2+2, \dots , N/2-1$, the grid points.

\begin{figure}
\begin{center}
\includegraphics[width=9cm]{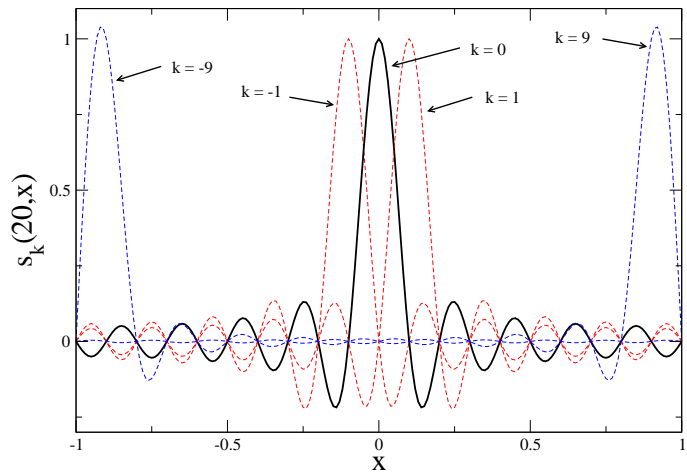}
\caption{LSF for $N=20$ and different values of $k$ (color online).}
\bigskip
\label{FIG1}
\end{center}
\end{figure}

We then define a set of $N-1$ LSF
\beq
s_k(h,N,x) \equiv  \frac{1}{2 N}
\left\{ \frac{\sin \left( \left(1+\frac{1}{2N}\right) \ \frac{\pi}{h} (x-k h) \right)}{
\sin \left( \frac{\pi}{2 N h} (x-k h) \right) }
-\frac{\cos\left(\left(1+\frac{1}{2N}\right) \ \frac{\pi}{h} (x+k h)\right)}{
\cos \left(\frac{\pi}{2 N h} (x+k h)\right)} \right\} \ .
\label{sincls}
\eeq
Fig.~\ref{FIG1} shows $5$ of these functions for $N = 20$.

It is not difficult to prove that the $s_k$ are orthogonal
\beq
\int_{-L}^L s_k(h,N,x) s_j(h,N,x) dx = h \ \delta_{kj} \ .
\eeq
and satisfy
\beq
s_k(h,N,y_j) = \delta_{kj} \ ,
\eeq
properties that are also shared by the sinc functions.

\begin{figure}
\begin{center}
\includegraphics[width=13cm]{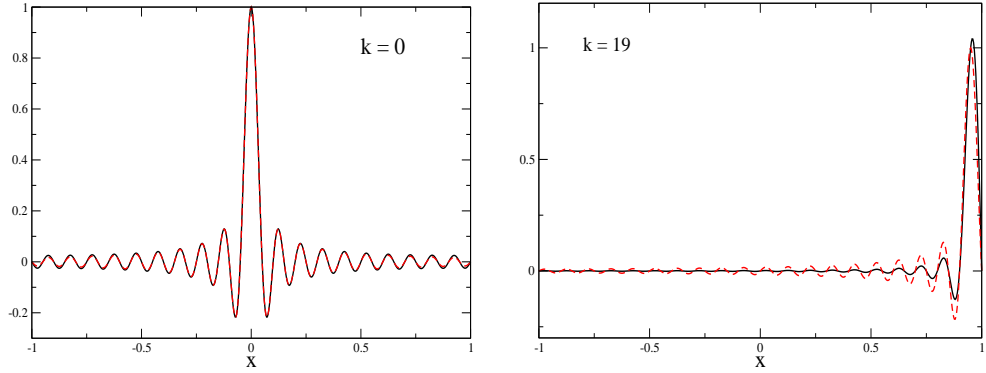}
\caption{Comparison between the sinc functions and  little sinc functions corresponding to $N=40$ (color online). }
\bigskip
\label{FIG2}
\end{center}
\end{figure}

Therefore, it is not surprising that the LSF become the standard sinc functions
when $N \rightarrow \infty$ and $h$ is held constant in eq.~(\ref{sincls}):
\beq
\lim_{N\rightarrow \infty} s_k(h,N,x) = \frac{\sin (\pi (x-k h)/h)}{\pi (x-k h)/h)} \equiv S_k(h,x) \ .
\label{sinc_asym}
\eeq
This property justifies the name of {\sl little sinc functions}.

Fig.~\ref{FIG2} shows the LSF for $N = 40$ and $L=1$ together with the corresponding sinc
functions. Differences between both kind of functions are appreciable only in the right plot,
corresponding to $k=19$. In this case the LSF is slightly larger than unity at the peak and
its oscillations die out faster.

The LSF share some properties with the sinc functions; for example, we can approximate a
function $f(x)$ on the interval $(-L,L)$ as
\beq
f(x) \approx \sum_{k=-N/2+1}^{N/2-1} f(x_k) \ s_k(h, N,x) \ ,\label{f(x)_LSF}
\eeq
where $x_k\equiv k h$.
Similarly one can express the derivatives of LSF in terms of LSF as
\beq
\frac{d s_k(h,N,x)}{dx} \approx \sum_j \left. \frac{d s_k(h,N,x)}{dx}\right|_{x=x_j} \ s_j(h,N,x) \equiv \sum_j c_{kj}^{(1)} \ s_j(h,N,x) \\
\frac{d^2 s_k(h,N,x)}{dx^2} \approx \sum_j  \left. \frac{d^2 s_k(h,N,x)}{dx^2}\right|_{x=x_j} \ s_j(h,N,x) \equiv \sum_j c_{kj}^{(2)} \ s_j(h,N,x)  \ ,
\label{der2}
\eeq
where the $c_{kj}^{(r)}$ are the counterpart of the coefficients shown above for the sinc
functions.

An explicit calculation yields
\beq
c_{jj}^{(1)} &=& \frac{\pi}{4 L}   \tan \left(\frac{j \pi }{N}\right) \\
c_{kj}^{(1)} &=& (-1)^{k-j} \frac{\pi}{4 L} \left(\cot \left(\frac{(j-k) \pi }{2 N }\right)+
\tan \left(\frac{(j+k) \pi }{2 N}\right)\right)
\label{c1}
\eeq
and
\beq
c_{jj}^{(2)} &=& -\frac{\pi ^2}{24 L^2} \left(1+2 N^2-3 \sec ^2\left(\frac{j \pi }{N}\right)\right) \\
c_{kj}^{(2)} &=& -(-1)^{j-k}  \frac{\pi^2}{8 L^2} \
\frac{\cos \left(\frac{j \pi }{N}\right) \cos\left(\frac{k \pi }{N}\right)}{
\cos^2\left(\frac{\pi}{2 N} (j+k)\right) \sin^2\left(\frac{\pi }{2 N} (j-k)\right)} \ .
\label{c2}
\eeq

Because of eq.~(\ref{sinc_asym}) these matrices reduce to the usual sinc
expressions given in the preceding section when $N \rightarrow \infty$.

It is straightforward to generalize from eq.~(\ref{sincls}), defined in $(-L, L)$, to
an arbitrary interval $(a,b)$:
\beq
\tilde{s}_k\left(h,N, x\right) \equiv s_k\left(h,N, \frac{2 L}{b-a} \
\left(x - \frac{a+b}{2}\right)\right) \ .
\eeq
In this case the points of the grid are given by
\beq
x_k = (b-a) \frac{k}{N-2} + \frac{a+b}{2} \ .
\eeq

In order to apply sinc collocation on a finite domain one commonly maps the real line
onto a finite interval~\cite{Ste93} by means of the conformal transformation
\beq
\phi(z) = \log \left(\frac{z-a}{b-z}\right) \ .
\label{map}
\eeq
This map carries the eye-shaped region
\beq
D_E = \left\{ z = x+ i y : \left|arg \left(\frac{z-a}{b-z}\right) \right| < d \leq \frac{\pi}{2}  \right\}
\eeq
into the infinite strip
\beq
D_S = \left\{ w = u + i v : \left|v \right| < d \leq \frac{\pi}{2}  \right\} \ .
\eeq

Under the inverse transformation, $z=\phi^{-1}(w)$ the points of the uniform grid on the real
axis, given by $u_k = k h$, are mapped onto the non uniform grid defined by the points
$\overline{x}_k = (a+b e^{k h})/(1+e^{kh})$.

In this case the sinc functions are mapped onto
\beq
\overline{S}_k(h,x) \equiv \frac{\sin\left(\pi (\phi(x)-k h)/h\right)}{\pi (\phi(x)-k h)/h} \ ,
\eeq
which equals unity at $x=\overline{x}_k$. Consequently, it is possible to approximate a function
in the interval $(a,b)$ as
\beq
f(x) \approx \overline{f}(x) = \sum_{k=-N}^{+N} f(\overline{x}_k) \ \overline{S}_k(h,x) \ .
\eeq

\begin{figure}
\begin{center}
\includegraphics[width=9cm]{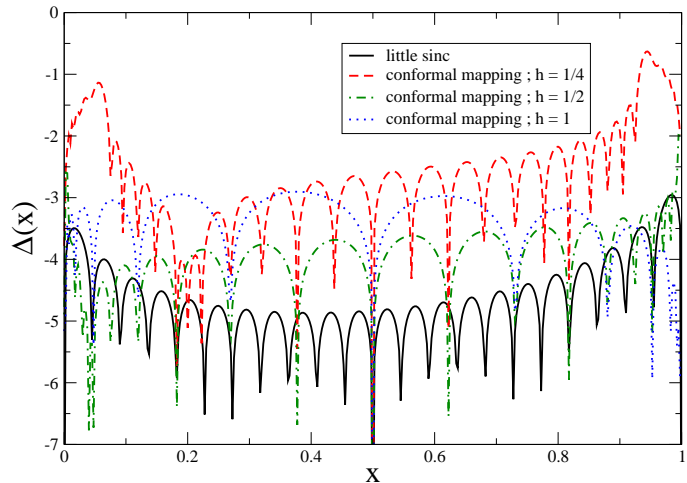}
\caption{Error in the interpolation of $f(x)$ in eq.(\ref{eq:steng}) using $21$ functions (color online).}
\bigskip
\label{FIG_err_steng}
\end{center}
\end{figure}

We can test the performance of the LSF on an example selected from Ref.~\cite{Nara02}:
\beq
f(x) = 2 x^2+x-3 x^3
\label{eq:steng}
\eeq
where $0 \leq x \leq 1$.  Fig.~\ref{FIG_err_steng} compares the logarithmic error
$\Delta(x) \equiv \log_{10} \left| f(x)-\overline{f}(x)\right|$ for both kind of functions.
The solid curve corresponds to $21$ LSF, whereas the other three curves correspond
to the same number of conformally mapped sinc functions with spacing $h = 1/4, 1/2$, and $1$
respectively. The LSF produce smaller errors and offer the advantage of a uniform grid.

Stenger originally introduced sinc methods for the numerical solution of differential
equations~\cite{Ste79}. Sinc-Galerkin and sinc collocation methods are particularly
useful in dealing with these problems since they converge exponentially even in the presence of
boundary singularities. It is not our purpose to generalize all the known mathematical
results from the sinc functions to the LSF; however, we assume that both kind of functions
share similar properties and simply compare our LSF results with those provided by the
conformally mapped sinc functions.

We consider the example $4.1$ of Ref.~\cite{Lybeck96}; that is to say, the inhomogeneous
differential equation~\footnote{Notice a typo in \cite{Lybeck96}, where $69 x$ must read $62 x$.}
\beq
-u''(x)+u'(x)+u(x)= \left(\frac{4}{25}\right)^2 \ \left( x^4 - 2 x^3 - 29 x^2 + 62 x
+ 38 \right) \label{eq:Lybeck}
\eeq
with the boundary conditions $u(-1) = u(4) = 0$.
The exact solution to this equation is
\beq
u_{exact}(x) =  \left(\frac{4}{25}\right)^2 \ (x+1)^2 \ (x-4)^2 \ .
\eeq
We look for a numerical solution in terms of our LSF as
\beq
u(x) \approx u_{ls}(x) = \sum_{k=-N/2+1}^{N/2-1} u_k \ s_{k}(h,N,x-3/2) \ .
\eeq

\begin{figure}
\begin{center}
\includegraphics[width=9cm]{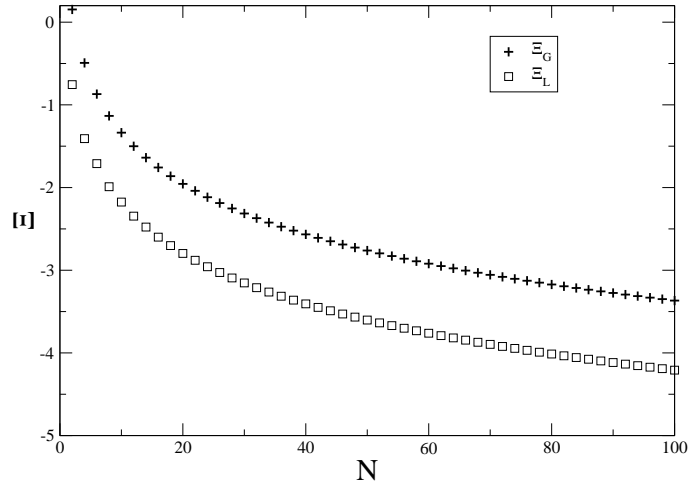}
\caption{Global and local errors (\ref{xig}) and (\ref{xil}), respectively, for
the solution of eq.~(\ref{eq:Lybeck}) in terms of $N-1$  LSF.}
\bigskip
\label{FIG_Lybeck_1}
\end{center}
\end{figure}

Fig.~\ref{FIG_Lybeck_1} shows global and local errors defined respectively as
\beq
\Xi_G(N) &\equiv& \log_{10} \left| \int_{-1}^{4} (u_{exact}(x)-u_{ls}(x))^2 \ dx \right| \nonumber \\
&=& \log_{10} \left|\int_{-1}^4 u_{exact}^2(x) \ dx - h \sum_{k=-N/2+1}^{N/2-1} u^2_k\right|
\label{xig}
\\
\Xi_L(N) &\equiv& \log_{10} \left| (u_{exact}(3/2) - u_{ls}(3/2) \right| \ .
\label{xil}
\eeq

Notice that at large $N$ the errors appear to decay exponentially. Unlike Ref.~\cite{Lybeck96} no
domain decomposition was required to solve this problem.

\begin{figure}
\begin{center}
\includegraphics[width=9cm]{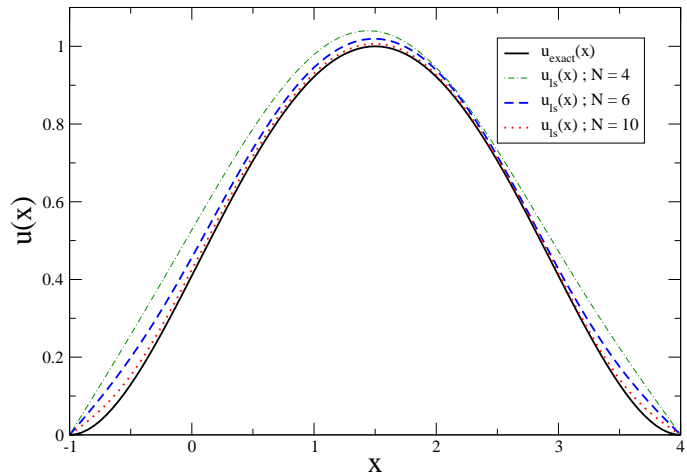}
\caption{Comparison between the exact solution $u_{true}(x)$ (solid line) and the
approximations obtained using LSF with $N = 4,6, 10$ (color online).}
\bigskip
\label{FIG_Lybeck_2}
\end{center}
\end{figure}


\section{The Schr\"odinger equation}
\label{sch}

As mentioned in the Introduction, sinc collocation methods have been used to obtain accurate
numerical solutions to the Schr\"odinger equation. In this section we wish to extend the
variational approach and the results of Ref.~\cite{Amore06} to our LSF and generalize the
method enclosing potentials with both bound and unbound states.  To simplify this
presentation we begin with potentials containing only bound states.

\subsection{Potentials containing only bound states}

We consider the one dimensional Schr\"odinger equation
\beq
- \frac{\hbar^2}{2m} \frac{d^2\psi_n(x)}{dx^2} + V(x) \psi_n(x) = E_n \psi_n(x)
\label{scheq}
\eeq
on an interval $(a,b)$, which can be either finite or infinite.
The wave functions $\psi_n(x)$ obey the boundary conditions $\psi_n(a) = \psi_n(b) = 0$
which grant that there will be only bound states.

If we express the wave functions in terms of our LSF, then
equations (\ref{f(x)_LSF}), (\ref{der2}) and (\ref{c2}) allow one to derive
the following matrix representation of the hamiltonian operator
\beq
H_{kl} = - \frac{\hbar^2}{2m} c_{kl}^{(2)} + \delta_{kl} V(k h)  \ .
\label{Hmat}
\eeq
This equation is similar to  eq.(8) of Ref.~\cite{Amore06}, except for the form of the
matrix $\mathbf{c}^{(2)}$. Once the spacing $h$ of the grid is fixed, then the
$N \times N$ matrix $\mathbf{H}_N$
for a manifold of $N$ LSF can be diagonalized. In this way one obtains an approximation to the
lower part of the spectrum, consisting of the first $N$ eigenvalues and wave functions
of the Schr\"odinger equation~(\ref{scheq}).

Going back to eq.~(\ref{Hmat}) we observe that the precision of the approximate results
obtained by diagonalization of the hamiltonian matrix depends crucially on both the number of
LSF as well as on the grid spacing. In fact, altough a small spacing can help to increase
the resolution, if the number of the sinc functions is not large enough, the approximation
will not be able to grasp the natural scale of the problem and the overall precision will
be poor. On the other hand, a large spacing will certainly provide poor results, because
the details of the problem will not be resolved.
On these grounds it is easy to convince oneself that it is likely to exist an optimal spacing,
for a given number of functions. Finding this optimal spacing will allow one to reach
sufficiently accurate results with a relatively small number of grid points.

It was found earlier~\cite{Amore06} that an optimal spacing can be obtained by straightforward
application of the principle of minimal sensitivity (PMS) to the trace of the $N\times N$
hamiltonian matrix ${\cal T}_N(h) = Tr\left[ \mathbf{H}_N\right]$.
In fact, given that the trace of a hamiltonian is invariant under unitary transformations,
and that in the limit $h \rightarrow 0$ it will therefore be independent of
$h$, then the optimal spacing may be given by the solution of the
equation~\footnote{Notice that such property was invoked earlier in Ref.~\cite{Am_1} when using a
basis of harmonic oscillator wave functions depending on an arbitrary scale parameter}
\beq
\frac{d}{dh} {\cal T}_N (h) = 0.
\label{pmseq}
\eeq

Therefore the optimal value of $h$ is found by numerically solving a single algebraic equation,
a modest computational task; thus, the interval length $L$ appearing in
LSF is treated as a variational parameter.

To test the performance of our method we apply it to the first example in Ref.~\cite{Baye86},
i.e the harmonic oscillator with $\hbar = 1$, $m=1/2$ and $\omega = 2$. Using a cartesian
mesh, Baye and Heenen reported errors smaller than
$10^{-3}$ for the first three eigenvalues and $N=10$.
On the other hand, our LSF approach with $N=10$ (which corresponds to $9$ sinc functions),
and a grid spacing optimized according to eq.~(\ref{pmseq}), yields errors
of $-4.86 \times 10^{-6}$, $1.2 \times 10^{-4}$ and $-1.6 \times 10^{-3}$ for the
same cases.

The authors of Ref.~\cite{Baye86} also observe that for $N=50$, the high eigenvalues
become very sensitive to the value of $h$, and that the variation of the $30th$ eigenvalue
with respect to $h$ presents a marked minimum around $h = 0.35$. It is remarkable that the
PMS condition for $N=50$ yields $L_{PMS} = 8.93$, corresponding to
$h_{PMS} = 2L_{PMS}/N \approx 0.357$, which is extremely close to the value quoted
by Baye and Heenen. It is worth noticing that while the optimal value of $h$
quoted by Baye and Heenen is the result of an empirical observation, the almost identical
value of $h$ given by the PMS is just the numerical solution of the algebraic
equation~(\ref{pmseq}), which requires negligible computer time.

\begin{figure}
\begin{center}
\includegraphics[width=9cm]{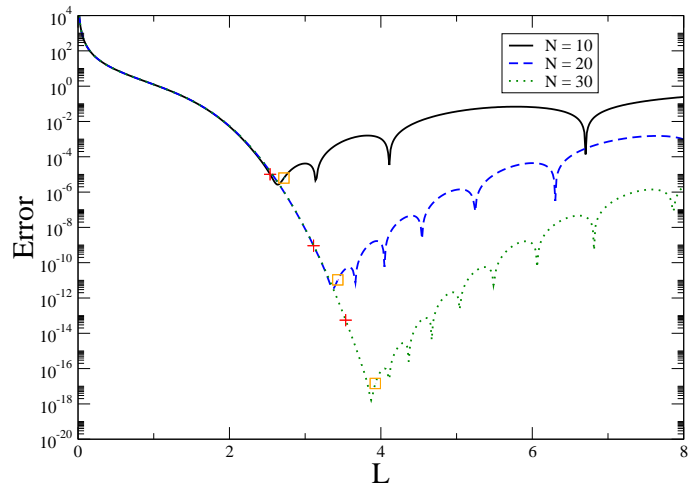}
\caption{Error in the ground--state energy of the potential (\ref{anharm}) using
$N = 10$, $20$ and $30$, as a function of the parameter $L$. The plus and square symbols
correspond to the PMS and modified PMS conditions, respectively (color online).}
\bigskip
\label{FIG_PMS}
\end{center}
\end{figure}

\begin{figure}
\begin{center}
\includegraphics[width=9cm]{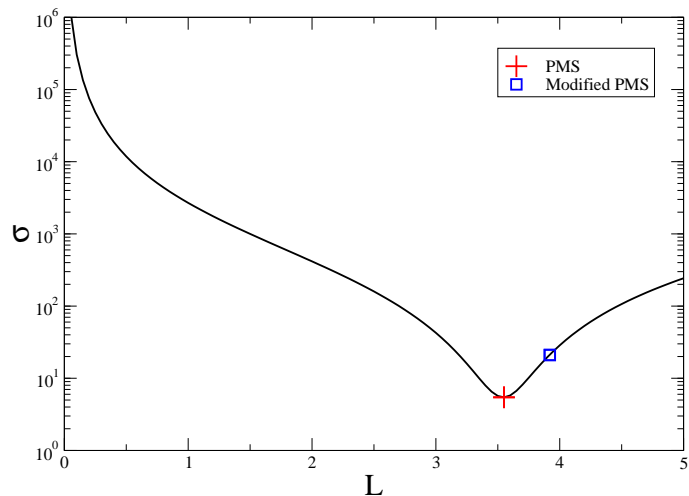}
\caption{Global error $\sigma \equiv \sum_{n=0}^{18}|E_n^{(20)}-E_n^{(60)}|$ for the potential
$V(x) = x^{4}/4$ (color online). }
\bigskip
\label{optimal}
\end{center}
\end{figure}

As a second example of application of the PMS to the LSF collocation method we consider
the anharmonic potential
\beq
V(x) = x^2 + x^4
\label{anharm}
\eeq
and we assume that $\hbar = 1$ and $m=1/2$ in the Sch\"rodinger equation.
This example was also studied by Baye and Heenen~\cite{Baye86} who obtained the optimal
values $h=0.55$ for $N=10$, and $h=0.2$ for $N=50$ using a cartesian mesh,
which is somehow related to the LSF.
Fig.~\ref{FIG_PMS} shows the error $|E_0^{exact}-E_0^{approx}|$ as a function of the
parameter $L$ for three different values of $N$ (remember that the number of LSF in the
expansion is $N-1$).
The plus symbols in the plot correspond to the predictions of the PMS condition,
which generally fall close to the minimum of the curve, while the square symbols correspond
to the solutions of a sort of empirical PMS condition, obtained by minimizing the modified
trace
\beq
\tilde{\cal T}_N(h) = {\cal T}_N(h) - \frac{N-2}{2 N} \sum_k V(x_k) \ .
\label{modPMS}
\eeq
with respect to $L$. The better behavior of the modified PMS condition for just one state
(as in Fig.~\ref{FIG_PMS}) should not confuse the reader. It must be kept in mind that the
PMS minimizes a sort of global error for all the states in the chosen manifold. In order to
appreciate this point clearly, Fig.~\ref{optimal} shows the global error
$\sigma = \sum_{n=0}^{N-1} | E_n^{(N)}-E_n^{(exact)}|$ as a function of
$L$ for the potential $x^4/4$ and $N = 20$. Notice that in this case the PMS condition
yields the minimal error. The exact energies $E_n^{(exact)}$ were simply chosen to be those
given by the method at higher order: $E_n^{(60)}$.

The accuracy of present calculations is greater than that obtained earlier
for the same problem~\cite{Baye86}, where the authors report errors of the order of $10^{-5}$
and $10^{-12}$ for $N=10$ and $N=50$, respectively.
Fig.~\ref{FIG_PMS} shows that the curves with different values of $N$ overlap in the
region of small $L$, which suggests that the approach may not be taking into account the
large--$L$ region correctly.

\subsection{Potentials containing bound and unbound states}

In what follows we try and show that the LSF are suitable for the treatment of potentials with
bound and unbound states. The application of the PMS condition to potentials with both discrete
and continuous spectra is not straightforward. The potentials treated in the preceding section
increase with the coordinate and therefore the matrix representation of those potentials
increase with $L$. Since the matrix representation of the kinetic energy decreases with $L$
then there is a minimum in the trace of the Hamiltonian matrix and the PMS gives a balance
between the traces of the kinetic and potential energies. That minimum provides the natural
length scale for the application of the method. If the potential energy tends to a
finite constant value as the coordinate increases, then there may not be a minimum and the
PMS will not yield the length scale for the application of present approach.

In order to overcome that problem we substitute a potential $\tilde{V}(x)$
into the Schr\"odinger equation that behaves exactly as the original potential $V(x)$
in the relevant coordinate region and increases to infinity at large distances.
The error introduced by such potential substitute will be negligible if the
difference between the original and substitute potentials is appreciable only where the
wavefunctions are expected to be vanishing small. This substitution removes the continuous
spectrum but should not affect the discrete one too much. The advantage is that
we are thus able to apply the PMS condition exactly as in the preceding subsection.
The potential substitute is introduced with the sole purpose of obtaining the length
scale and we diagonalize the correct hamiltonian matrix.

We will illustrate our procedure on the Morse potential already treated earlier
by means of the Lagrange mesh method~\cite{BGV02}:
\beq
V(r) = D \ \left[ e^{-2 a (r-r_e) } - 2 e^{- a (r-r_e)}  \right] \ ,
\eeq
where $D= 0.10262$, $r_e = 2$, $a= 0.72$, $2 m = 1836$ and $\hbar=1$.
In the case of states with nonzero angular momentum we should add the centrifugal
potential $\frac{\hbar^2 l (l+1)}{2 m r^2}$, $l=0, 1, 2, \dots$.

The potential substitute $\tilde{V}(r)$ is arbitrary; we can, for example, choose it to be
the Taylor expansion of $V(r)$ around a given point, and truncate it at a sufficiently
large order to be accurate enough at small $r$, and at the same time to satisfy
$\lim_{r\rightarrow \infty } \tilde{V}(r) = +\infty$.
For example, in present particular case we can choose:
\beq
\tilde{V}(r) = \sum_{n=0}^{20} \frac{1}{n!} \ \left.\frac{d^nV}{dr^n}\right|_{r=10} (r-10)^n \ .
\eeq

Another difficulty to take into consideration is that our LSF are defined
in $(-L,L)$ whereas the radial coordinate is defined in $(0,+\infty)$.
The obvious solution to this apparent problem would be using the form of Ref.~\cite{Baye95},
where the left boundary of the interval is $0$; on the other hand, such choice would
be equivalent to a shift by a proper amount of the potential, thus bringing the boundary
condition on the left point to coincide with the left point of the LSF. Despite its
simplicity, this procedure is not optimal and generally does not provide the best
results. We have found out that a more convenient strategy consists of keeping our
LSF unchanged and shifting the coordinate by given amount as
$V(r) \rightarrow V(r + \overline{r})$, where
$\overline{r}$ is typically close to the minimum of the potential (the same shift should be
applied to the centrifugal energy when necessary). The PMS applied
to the shifted hamiltonian will thus provide the optimal scale for the application
of the LSF collocation method.

One advantage of that procedure is that it maximizes the sampling of the
classically allowed region, where the bound state wave functions exhibit marked
nonzero contributions. Of course, in order to take into account the boundary
condition at $r=0$, the PMS length scale has to be smaller than $\overline{r}$.

\begin{table}
\caption{\label{tab1} Errors $\epsilon_n \equiv E_n^{approx}-E_n^{exact}$ for
some states of the Morse potential. Powers of ten are indicated between square brackets.}
\begin{ruledtabular}
\begin{tabular}{cccccc}
$l$ & $n$  & $N$ & Ref.\cite{BGV02}  & PMS & $h_{PMS}$\\
\hline
$0$ & $0$ & $20$ & $3.8 [-7]$   & $-3.3 [-10]$ & $0.223$ \\
    &     & $40$ & $< 1 [-14]$  & $1.8 [-20]$  & $0.134$ \\
    & $5$ & $20$ & $4.0 [-3]$   & $2.1 [-6]$  & $0.223$ \\
    &     & $40$ & $1.1 [-9]$   & $2.5 [-14]$  & $0.134$ \\
$1$ & $0$ & $20$ & $4.1 [-7]$   & $-3.6 [-10]$ & $0.223$ \\
    &     & $40$ & $< 1 [-14]$  & $-1.6 [-20]$  & $0.134$ \\
    & $5$ & $20$ & $4.4 [-3]$   & $1.5 [-6]$  & $0.223$ \\
    &     & $40$ & $1.1 [-9]$   & $2.9 [-14]$  & $0.134$ \\
$2$ & $0$ & $20$ & $4.5 [-7]$   & $-4.1 [-10]$ & $0.223$ \\
    &     & $40$ & $< 1 [-14]$  & $1.2 [-20]$  & $0.134$ \\
    & $5$ & $20$ & $4.7 [-3]$   & $2.0 [-7]$  & $0.223$ \\
    &     & $40$ & $1.1 [-9]$   & $3.9 [-14]$  & $0.134$ \\
\end{tabular}
\end{ruledtabular}
\bigskip\bigskip
\end{table}

\begin{figure}
\begin{center}
\includegraphics[width=9cm]{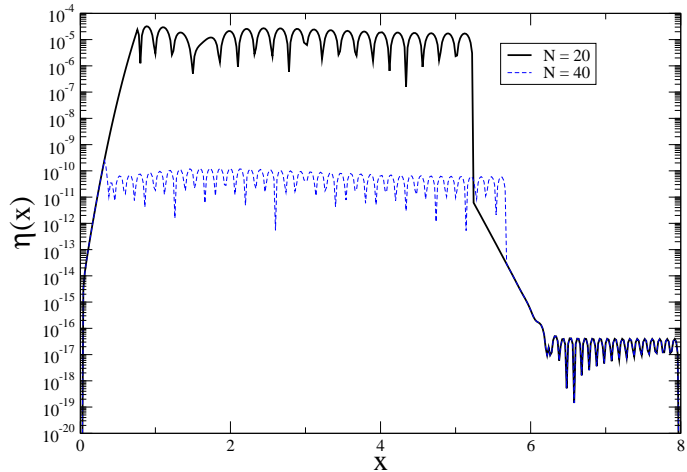}
\caption{Error $\eta_N(x) = | \psi^{(N)}(x)-\psi^{(80)}(x) |$ as a function of $x$
for the ground state of the Morse potential (color online).}
\bigskip
\label{FIG_eta}
\end{center}
\end{figure}

Table~\ref{tab1} shows the errors $\epsilon_n \equiv E_n^{approx}-E_n^{exact}$ for
the s, p, and d states of the Morse potential with $n=0$ and $n=5$. It compares
present results with those of Baye et al~\cite{BGV02}. The last column of
Table~\ref{tab1} displays the PMS optimal values of the grid spacing.

Fig.~\ref{FIG_eta} shows the local error $\eta_N(x) = | \psi^{(N)}(x)-\psi^{(80)}(x) |$ for
the ground state of the Morse potential and approximations $N=20$ and $N=40$,
where we have chosen $\overline{r}=3$. We assume that the approximation of order $N=80$
is sufficiently close to the exact wavefunction.  Our results are more accurate than those
of Baye et al~\cite{BGV02} who essentially considered the average of $\eta(x)$ over a
chosen region.  The sharp drop of the error beyond a certain value of $r$, clearly
noticeable in Fig.~\ref{FIG_eta}, is due to the fact that our LSF reproduce the wave function
only in a finite region outside which $\eta(x)$ is merely the value of $\psi^{(80)}(x)$.
It is worth noticing that $\eta_N(x)$ is almost uniform in the region covered by the LSF.

We have also applied our method to the one--dimensional Morse potential considered by
Wei~\cite{Wei00}:
\beq
V(x) = D \left[e^{-2 \alpha x} - 2 e^{-\alpha x} + 1 \right],
\eeq
where $-\infty<x<\infty$, $D = 0.0224$, $\alpha = 0.9374$, $m = 119406$ and $\hbar=1$. This problem can be solved
exactly and the energies are given by \cite{Marcelo}
\beq
E_n = \hbar \omega \left[  n + \frac{1}{2} - \frac{\hbar \omega}{4D} \left( n+\frac{1}{2}\right)^2  \right]
\eeq
where $\omega = \sqrt{2 D/m}\ \alpha$. As before we can improve our
results by conveniently shifting the potential on the $x$-axis from $V(x)$ to
$V(x+\overline{x})$. When $N = 20$ and assuming $\overline{x}=0$ we have found
that the first--excited--state eigenvalue is reproduced with an accuracy of about
$8 \times 10^{-14}$,
which is even better than the accuracy obtained by Wei with
$N=64$~\footnote{Notice that Table~I of Ref.~\cite{Wei00} omits the the ground state
of the model.}. When $N = 40$ the error of our method is just $8.1 \times 10^{-21}$
for the ground state.

\section{Conclusions}
\label{concl}

In this paper we have introduced a new class of sinc-like functions, which we name
``little sinc functions'' (LSF), which share some properties of the usual sinc
functions, but are defined on finite intervals.
We have shown that the LSF collocation method provides accurate approximations for
a wide class of problems when it is supplemented by the Principle of Minimal
Sensitivity (PMS) that provides an optimal grid spacing. In particular, we have
applied the LSF to the solution of the
Schr\"odinger equation with only bound states and with mixed discrete and continuous
spectra. We have chosen benchmark models treated earlier by other authors and in all
the cases we obtained more accurate results. It seems that the LSF collocation method
is an interesting alternative algorithm for solving many mathematical and physical problems.

\begin{acknowledgments}
P.A. acknowledges support of Conacyt grant no. C01-40633/A-1.
\end{acknowledgments}

\end{document}